\documentstyle[epsfig, subfigure]{mn}

\newcommand{\pre}{Physical Review E}

\newcommand{\aap}{A\&A}
\newcommand{\mnras}{MNRAS}
\newcommand{\apj}{ApJ}
\newcommand{\apjl}{ApJ}

\newcommand{\memsai}{MmSAIt}

\newcommand{\nat}{Nature}
\newcommand{\aapr}{Astronomy \& Astrophysics Rev.}
\newcommand{\araa}{Annual Rev. of Astronomy \& Astrophysics}

\newcommand{\jcap}{Journal of Cosmology and Astroparticle Physics}

\newcommand{\jgr}{Journal of Geophysical Research}

\def\ltapprox{\raise 2pt \hbox {$<$} \kern-1.1em \lower 5pt \hbox {$\approx$}}
\def\ltsim{\raise 2pt \hbox {$<$} \kern-1.1em \lower 4pt \hbox {$\sim$}}
\def\gtsim{\raise 2pt \hbox {$>$} \kern-1.1em \lower 4pt \hbox {$\sim$}}

\title{Stochastic reacceleration of relativistic electrons by  
turbulent reconnection: a mechanism for cluster-scale radio emission ?}

\author[G. Brunetti, A. Lazarian]
      {G. Brunetti,$^1$ 
       A. Lazarian$^2$\\
       $^1$ INAF/Istituto di Radioastronomia, via Gobetti 101,
       I--40129 Bologna, Italy \\
       $^2$ Department of Astronomy, University of Wisconsin at Madison,
       5534 Sterling Hall, 475 North Charter Street, Madison,
       WI 53706, USA\\
}

\begin{document}
\maketitle

\begin{abstract}
In this paper we investigate a situation where relativistic
particles are reaccelerated diffusing across regions of 
reconnection and magnetic dynamo in super-Alfvenic,
incompressible large-scale turbulence. 
We present an exploratory study of this mechanism in the
intra-cluster-medium (ICM). In view of large-scale turbulence in the
ICM we adopt a reconnection scheme that is based on turbulent
reconnection and MHD turbulence.
In this case particles are accelerated and decelerated 
in a systematic way in reconnecting and magnetic-dynamo 
regions, respectively, and on longer time-scales undergo a stochastic process 
diffusing across these sites (similar to second-order Fermi).
Our study extends on larger scales numerical studies 
that focused on the acceleration in and around turbulent reconnecting regions.
We suggest that this mechanism may play a role in the reacceleration
of relativistic electrons in galaxy clusters providing a new physical
scenario to explain the origin of cluster-scale diffuse radio emission.
Indeed differently from current turbulent reacceleration models proposed
for example for radio halos
this mechanism is based on the effect of large-scale incompressible
and super-Alfvenic turbulence. In this new model 
turbulence governs the interaction between 
relativistic particles and magnetic field lines that diffuse, reconnect
and are stretched in the turbulent ICM.
\end{abstract}


\maketitle

\section{Introduction}

Magnetic reconnection is a long-standing problem in astrophysics and
plasma physics. 

\noindent
The traditional Sweet-Parker model of
reconnection (Parker, 1957; Sweet, 1958)
is very slow in all the astrophysical settings, which induced to
searches of faster reconnection schemes.
Petschek (1964) or X point reconnection for years has been the major
contender for the mechanism for fast astrophysical reconnection. 
With plasma effects, e.g. Hall effect, included, the Petschek scheme 
looked promising to many researchers in the field (eg. Shay et al 1998, 
Birn et al 2001, Bhattacharjee 2004).

\noindent
In parallel a scheme that invokes 3D Alfvenic turbulence was suggested
in Lazarian \& Vishniac (1999, henceforth LV99). 
The process appealed to MHD turbulence properties and
predicted the independence of the reconnection rates on microphysics. 
Numerical simulations with externally induced turbulence supported
analytical predictions in LV99 (eg. Kowal et al. 2011, 2012).

\noindent
More recently, the Petschek scheme evolved into tearing
reconnection, which presents much more chaotic state of the 
magnetic field evolution (Loureiro et al 2007, Uzdensky et al. 2010).
Numerical simulations of tearing reconnection show a transition to
turbulent state, although the role of turbulence for the reconnection 
in these simulations is debated (e.g., Karimabadi \& Lazarian 2013 for
review). 
Other simulations are suggestive of fast, i.e. independent of resistivity, 
magnetic reconnection even in the case when turbulence is not induced
externally, but is generated by reconnection itself
(Beresnyak 2013, Oishi et al. 2015, Lazarian et al. 2015). 

\noindent
These two schemes, the turbulent and tearing one, present at the moment
two competing reconnection processes, which, as we discuss further 
in the paper, can be complementary and synergetic. 
In view of large-scale 
turbulence being present in the intra-cluster medium (ICM) 
(e.g., Brunetti \& Jones 2014 (henceforth BJ14), Brueggen \& Vazza 2015 
for reviews), within this paper we mostly focus on the turbulent 
magnetic reconnection scheme.

The main aim of this paper is to explore a scenario of particle
acceleration in the ICM where relativistic electrons 
are reaccelerated by the combined effect of super-Alfvenic turbulence 
and magnetic reconnection.
As already mentioned 
turbulence and magnetic reconnection are interconnected aspects 
and reconnection can be fast in the presence of turbulence 
(LV99, Lazarian et al 2015 for review).
In the presence of 
super-Alfvenic turbulence relativistic particles can 
diffuse across a complex network of regions of reconnection
and magnetic field dynamo, where they are systematically
accelerated or decelerated.
This situation can be described as a combination of a first-order
Fermi acceleration and a second order Fermi acceleration 
due to spatial random-walk across accelerating and
decelerating regions.
Under the assumption of {\it fast} diffusion from reconnection regions 
this chain of mechanisms becomes a second-order 
acceleration process and is the main focus of this paper.

Galaxy clusters are potential sites for this mechanism 
because super-Alfvenic turbulence is generated in the ICM.
Evidence for relativistic electrons 
and magnetic fields in the ICM is routinely obtained
from a variety of radio observations that detect diffuse synchrotron
radiation from the ICM in the form of radio halos and 
relics (Feretti et al 2012, BJ14 for reviews).
Giant radio halos and relics
are found in disturbed clusters, suggesting that
large scale motions that are generated during mergers provide
the energy reservoir for the mechanisms of {\it in situ} particle 
acceleration (BJ14 for review on acceleration mechanisms in the ICM).

\noindent
A popular scenario for the origin of radio halos is based on 
reacceleration induced by merger-driven turbulence
(Brunetti et al 2001, Petrosian 2001, Fujita et al. 2003,
Brunetti et al. 2004, BJ14 for review). In this model
it is usually assumed that 
large-scale compressive turbulence 
generated during cluster mergers is 
transported at smaller scales via turbulent cascade and reaccelerates
seed electrons (Cassano \& Brunetti 2005, Brunetti \& Lazarian 2007).
This scenario has the potential to explain the observed 
properties of radio halos and it naturally explains their 
connection with mergers, still several challenges exist.
Primarily the challenge is to determine the efficiency of
particle acceleration that depends on the complex hierarchy of 
mechanisms that transport energy from large scale motions
to collisionless
particle-wave interactions at small-scales in the ICM (eg. Brunetti 2016 
for a recent discussion).
An additional challenge is that turbulent acceleration in the ICM
is a slow process that cannot accelerate particles directly 
from the thermal pool (eg. Petrosian \& East 2008). For this
reason turbulent acceleration models require 
a population of seed electrons that are already ultra-relativistic 
(e.g., $\gamma \geq 100$) and that are spatially distributed on
cluster scales (eg. Brunetti et al 2001,
Brunetti \& Lazarian 2011a, Pinzke et al 2015).

\noindent
More recently numerical (fluid) simulations suggested that the
compressive part of turbulence in the ICM is partially dissipated 
into weak shocks (see Miniati 2015, Porter et al 2015),
implying that less energy is available for the 
mechanisms that are generally invoked in turbulent reacceleration models.

\noindent
In order to overcome this last challenge we exploit our mechanism that is 
based on incompressible turbulence.
We show that this mechanism may play a role in the reacceleration of 
relativistic electrons in the ICM under reasonable assumptions 
on the effective particle mean-free path.

In Sect.2 we discuss turbulent reconnection in super-Alfvenic turbulence.
In Sect.3 we explain the acceleration mechanism and in Sect.4 we apply
it to the case of cluster-scale radio emission in galaxy clusters.
In Sect.5 we provide a discussion on CRs acceleration in the ICM and
on the comparison between our model and other turbulent
reacceleration models. We summarize results in Sect.6.

\section{Turbulence and magnetic reconnection}

The ICM is a {\it weakly collisional} high-beta plasma and is expected to 
be turbulent at some level (BJ14, Bruggen \& Vazza
2015 for reviews), the presence of instabilities in this plasma 
would make turbulence similar to MHD one (Santos-Lima et al 2015 
and references therein).
Natural drivers of ICM turbulence are the motions 
gravitationally driven by dark matter substructures 
that are generated in the ICM as a consequence of the hierarchical process of 
formation of galaxy clusters.
As a consequence of these motions super-Alfvenic and sub-sonic 
turbulence should be driven in the ICM, 
although the way this turbulence is transported from large scale 
to small scale depends on details of ICM microphysics that are 
still poorly understood. 
It is believed that a hierarchy of processes in the ICM
convert turbulent energy into 
non-linear amplification of magnetic fields, and particle 
heating and acceleration (eg. Ryu et al 2008, BJ14,
Miniati \& Beresnyak 2015).

\noindent
Magnetic reconnection and dynamo are part of the turbulent cascade 
(eg. Lalescu et al 2015) and play a role in this complex hierarchy of 
mechanisms.
Specifically, the model of turbulent reconnection by  
LV99 provides a natural extension of the 
classical Sweet-Parker model in the presence of 
turbulent motions.

\noindent
Macroscopically the reconnection speed is constrained by the possibility to
eject the plasma from the reconnection layer.
It gives a maximum reconnection speed on macroscopic scales :

\begin{equation}
V_{rec} \approx V_A {{\Delta}\over{l_X}}
\label{vrec}
\end{equation}

\noindent
where $\Delta$ is the thickness of the reconnection region (current sheet)
and $l_X$ is the astrophysical scale of the inflow associated with
a reconnection region.
In the classical Sweet-Parker model two regions
with uniform {\it laminar} magnetic fields are separated by a thin
current sheet, which
is determined by microscopic resistivity. Over the thickness of this
thin current sheet, $\Delta$, the
resistivity is important and the magnetic fields reconnect.
However in astrophysical situations the disparity of
scales between the inflow, $l_X$, and the outflow, $\Delta$, 
strongly limit the reconnection rate (from Eq.\ref{vrec}), implying
$V_{rec} \ll V_A$.

\noindent
One possibility to overcome this bottleneck is turbulence.
Turbulence changes the direction of magnetic field lines that
are subject to turbulent diffusivity.
Under these conditions the outflow $\Delta$ gets much thicker, 
being determined not by microphysics of resistivity, but by macroscopic 
field wandering in 3D (LV99).
LV99 derived the reconnection rate for 
sub-Alfvenic turbulence, this is $V_{rec} \sim (l_X/L_o)^{1/2} M_A^2 V_A$,
where $L_o$ and $M_A$ are the turbulent injection scale and Alfven Mach
number, respectively. 

\noindent
Results can be easily extended into the
super-Alfvenic regime (Lazarian et al 2015) that is typical of the
ICM.
An intuitive way to estimate $\Delta$ in super-Alfvenic turbulence
is to use the diffusion properties of magnetic field lines.
In this approach $\Delta$ is essentially the diffusion scale
covered by magnetic field lines within the time-period necessary to
expel the plasma from the reconnection region, $\tau_A \sim
l_X/V_A$.
It has been shown that field diffusion in these conditions is similar to
Richardson diffusion in hydro- motions (Eyink et al 2013). 
It implies $\Delta^2 \approx \epsilon \tau_A^3$, where 
$\epsilon \sim \delta V^3/L_o$ is the specific turbulent energy rate, 
from which one immediately gets $\Delta \sim l_A$.

\noindent
The other ingredient to determine a macroscopic constraint to the
reconnection speed in Eq.\ref{vrec}
is the inflow, astrophysical, scale 
of reconnection regions $l_X$.
Super-Alfvenic turbulence naturally generates regions of magnetic
field reversals everywhere in the plasma
with minimum scale of the order of the 
MHD scale $l_A = L_o M_A^{-3}$. At scales $l > l_A$ 
kinetic energy is in excess of magnetic energy inducing a 
continuous change of magnetic field topology, due to advection of field
lines by hydro- motions, and consequently quenching the reconnection
process. 
Under these conditions the dominant scale for the inflow in 
reconnection regions is expected to be $l_X \sim l_A$.

\noindent
As a consequence we may think of a situation 
of {\it fat} reconnection regions in super-Alfvenic turbulence, with
$\Delta \sim l_X \sim l_A$, where the 
macroscopic reconnection speed can approach the Alfven speed
(Eq.\ref{vrec}).

\noindent
In the following we shall adopt this configuration to calculate 
particle acceleration.

\section{Acceleration of relativistic particles in reconnection}

\subsection{Brief Overview on acceleration in turbulent reconnection}

Particle acceleration in reconnection regions has been
investigated in different environments and considering different
mechanisms (e.g., Lyutikov \& Blandford 2003, Lyubarsky 2003,
Drake et al 2006, 2010, 2013, Lazarian \& Opher 2009, 
Lazarian \& Desiati 2010, Kowal et al 2011, 
Giannios 2013, Sironi \& Spitkovsky 2014, Zank et al 2014).

\noindent
One mechanism has been proposed within turbulent reconnecting regions.
Turbulent reconnection results in shrinking of 3D magnetic bundles 
and the charged particles entrained over magnetic loops can be
accelerated by firts-order Fermi 
mechanism (de Gouveia dal Pino \& Lazarian 2005).
This mechanism has been used to model particle acceleration in
several environments, including gamma-ray bursts (Zhang \& Yan 2011),
low luminosity AGNs (Kadowaki et al 2015), microquasars (Khiali et
al 2015) and in the heliosphere (Lazarian \& Desiati 2010).
Interestingly, the compressibility is not necessary in this
mechanism \footnote{a generalised version of the mechanism in the presence of 
compression is discussed in Drury (2012)}.
The physics of the acceleration is easy to understand. 
Magnetic reconnection induces systematic shrinking of magnetic field lines on
scales $\leq l_A$ and particles entrained on these field lines are getting 
energy as the consequence of the Liouville theorem.

\noindent
Particle acceleration in turbulent reconnection sites has also
been investigated by numerical studies that combine MHD simulations and tracer
particles (eg., Kowal et al 2011,12).
The acceleration of particles trapped within converging reconnection regions 
is consistent with first-order Fermi mechanisms.

\noindent
In a general situation  
presumably particles interact with multiple reconnection 
layers and turbulence experiencing a hierarchy of processes at different
scales and energies. 
An exploratory investigation of this situation is the main focus of this
Section (Sect.3.2). 
However 
we anticipate that the basic picture deriving from our study does not
depend critically on the specific mechanism of particle acceleration 
within reconnecting regions (see Sect.3.3).

\subsection{Our approach}

Here we follow a simple approach 
to explore the general situation where relativistic
particles are reaccelerated diffusing on large-scale super-Alfvenic
turbulence and reconnection regions.
We proceed with the following assumptions:

\begin{itemize}
\item[{\it i})] large-scale
turbulence in the ICM behaves collisional and 
MHD provides a good guide as far as reconnection is considered;

\item[{\it ii})] ultra-relativistic
CRe are collisionless components interacting with 
fluctuations in the
MHD fluid;

\item[{\it iii})] we assume a scenario based on
turbulent reconnection, where diffusion of magnetic field lines 
governs the process of scattering of CRs;

\item[{\it iv})] we expressly investigate only the 
collisionless effects that involve the
interactions between large-scale magnetic field lines and CRe.
\end{itemize}

\noindent
It is difficult to establish to what extent these assumptions allow 
a solid description of the ICM.
The main point here is that we need reconnection events in terms of
(belonging to) MHD cascade.
However it is worth to note that even if turbulence in the
ICM is collisionless we may still expect that
the dynamics of the magnetic field lines can be described 
according to Alfvenic turbulence (Eyink et al
2011, Makwana et al 2015), thus supporting our basic assumptions. 

We start considering particles within reconnecting regions.
The increment in momentum of
particles after the interaction with shrinking magnetic field
lines can be estimated as :

\begin{equation}
\Delta p \simeq \Delta t \times {{dp}\over{dt}}
\label{deltap}
\end{equation}

\noindent
where the time-scale spent by particles in a reconnecting region
is :

\begin{equation}
\Delta t = 
\min \left\{ {{l_A}\over{V_A}},{{l_A^2}\over{D}} \right\}
\label{Dt}
\end{equation}

\noindent
where ${{l_A}\over{V_A}}$ is the turn-over time of eddies
(corresponding to {\it slow} diffusion limit)
and ${{l_A^2}\over{D}}$ is the particle diffusion time on scale $l_A$
(corresponding to {\it fast} diffusion, $D$ is the spatial diffusion 
coefficient of CRe), and 

\begin{equation}
{{dp}\over{dt}}
\sim
\phi {{V_A}\over{\lambda_{mfp}}} p
\label{dpdt}
\end{equation}

\noindent
where $\lambda_{mfp}$ is the particles mean free path (mfp)
and $\phi \sim 1$ 
accounts for pitch angle distribution and geometry of scatterings.

\noindent
However, in a turbulent fluid with stationary
turbulence the
reconnection process should be accompanied by the opposite process
of magnetic field generation, i.e. by turbulent dynamo. 
Naturally, this process results in a stretching of the magnetic
field lines that on scale $l_A$ occurs on a time-scale $l_A/V_A$
and that {\it statistically} compensates field diffusion on the same
scale (Beresnyak \& Lazarian 2015).
It results in a situation where energetic particles within dynamo regions 
cool by an amount that is similar (in absolute value) 
to that of the acceleration in reconnection regions.

\noindent
Therefore if particles can diffuse across these regions, we 
expect that they will undergo cycles of first order
acceleration in reconnection regions and cycles of cooling in dynamo
regions. 
This is the theoretical picture that we use to describe the
acceleration of particles diffusing in a turbulent
fluid on scales larger than $l_A$.
It differs from the traditional second order Fermi acceleration because 
the increments of the momentum of the CR in the acceleration and 
deceleration regions 
in principle can be comparable with the momentum of the CR.

In order to estimate the acceleration rate resulting 
from this complex mechanism we make the educated guess
that reconnecting and dynamo regions are statistically distributed 
everywhere in the astrophysical volume on typical scales $\sim l_A$.
This can be motivated by the fact that the generation of magnetic fields
and field reconnection occur indeed everywhere in the volume, 
although this assumption needs to be
evaluated more carefully in follow up studies.

\noindent
For $\Delta p \ll p$ in Eq.\ref{deltap}, particles interacting with
these regions will undergo
a random-walk process in the momentum space resembling a 
second-order Fermi mechanism. From Eqs.\ref{deltap}--\ref{dpdt} 
this condition implies 
$\lambda_{mfp} \gg \phi l_A$ (in the slow diffusion)
or $\lambda_{mfp} > l_A ( 3 \phi V_A/c)^{1/2}$ 
({\it fast} diffusion, assuming $D \sim {1 \over 3} c \lambda_{mfp}$). 
This also tells us that the condition $\Delta p \ll p$ in the ICM 
is possible only in the case of {\it fast} diffusion, because 
{\it slow} 
diffusion (Eq. \ref{Dt}) occurs for $\lambda_{mfp} \leq 3 V_A l_A/c
\ll l_A$.

In the following we will restrict to the case of fast diffusion and 
to situations where $\Delta p \ll p$.
Under these conditions particles undergo a diffusion 
process in the momentum space 
and retain a isotropic distribution of pitch angles, provided
that the pitch-angle scattering rate, $\tau_{sc} \sim \lambda_{mfp}/c$, 
is fast enough.
The distribution function of particles in the momentum space,
$f(p)$, evolves according to a isotropic Fokker-Planck equation
(e.g., Schlickeiser 2002):

\begin{eqnarray}
{{\partial f(p,t)}\over{\partial t}} =
{1 \over{p^2}} {{\partial  }\over{\partial p}} \left(
p^2 D_{pp} {{\partial f }\over{\partial p}}
- p^2 \sum \big| {{d p}\over{dt}} \big|_{loss} f(p,t) \right)
\nonumber\\
+ Q(p,t) - {{f(p,t)}\over{T_{esc}}}
\label{FP}
\end{eqnarray}

\noindent
where the diffusion coefficient in the particle momentum
space is (Eqs.\ref{deltap}--\ref{dpdt}) :

\begin{equation}
D_{pp} = \langle {{\Delta p \Delta p}\over
{2 \Delta t}} \rangle \sim 3 \left( {{l_A}\over{\lambda_{mfp}}}
\right)^2 {{V_A^2}\over{\lambda_{mfp} c}}
p^2
\label{accelerationrate}
\end{equation}

\noindent
implying a reacceleration time $\tau_{acc} \sim p^2/(4 D_{pp})$, and
where 
$\sum \big| {{d p}\over{dt}} \big|_{loss}$ accounts for
the energy losses in the ICM (synchrotron, ICS, Coulomb, ..), 
$Q$ accounts for injection of new
particles, and $f/T_{esc}$ accounts for diffusion/escape of particles
from the region ($T_{esc}$ is the escaping time).

\noindent
Solutions of Eq.\ref{FP} in the case $D_{pp}\propto p^2$ 
are calculated in numerous papers (eg Schlickeiser 1984, Mertsch 2011,
for numerical solutions Donnert \& Brunetti 2014 and references therein).
We anticipated that $D_{pp} \propto p^2$ also results from 
reacceleration by compressible turbulence (both Transit-Time-Damping and
stochastic compressions) in the ICM 
and consequently the properties of the
spectra of reaccelerated electrons 
expected in our model will be similar to those calculated in 
other studies (BL07, Brunetti \& Lazarian 2011a, ZuHone et al
2013, Donnert \& Brunetti 2014; see Sects. 4 and 5).

The acceleration rate from Eq.\ref{accelerationrate}
depends on the particles mfp that however
is poorly known. 
In super-Alfvenic turbulence hydro- motions change directions
of magnetic field 
lines on scales $\geq l_A$. 
This configuration automatically defines a maximum effective 
mfp of particles, $\lambda_{mfp} \sim l_A$, because particles
travelling in tangled fields change directions on this scale 
preserving the adiabatic invariant.
In addition magnetic field fluctuations in MHD turbulence induce 
particles pitch-angle scattering via resonant interaction and
affect the motion of particles.
This is different from the previous interaction because in this
case particles experience pitch-angle scattering with respect to the 
local field direction and do not preserve adiabatic invariant.
The parallel mfp from this process is :

\begin{equation}
\lambda_{mfp} =
{{3 c} \over 4}
\int_0^1
d\mu {{\left( 1 - \mu^2 \right)^2 }\over{
D_{\mu \mu}
}}
\label{mfpdmumu}
\end{equation}

\noindent
where $D_{\mu \mu}$ is the pitch-angle diffusion coefficient.
In the case of incompressible MHD turbulence on scales $< l_A$
resonant pitch-angle scattering is dominated by transit-time-damping
with pseudo-Alfven modes (Yan \& Lazarian 2008, Beresnyak et al 2011).
In the super-Alfvenic case the interaction is dominated by the largest
moving mirrors with $l \sim l_A$, and from Eqs 8-9 in
Yan \& Lazarian (2008) (in the limit $M_A \rightarrow 1$ 
and $L \rightarrow l_A$) we find $D_{\mu \mu} \sim {c \over{l_A}} F(\mu)$, 
where $F(\mu) \sim 1$ for transverse propagation 
and $F \ll 1$ in the quasi-parallel case.
In this case Eq.\ref{mfpdmumu} gives $\lambda_{mfp} \sim l_A$.

\noindent
In conclusion particles interacting with large-scale incompressible and
super-Alfvenic MHD turbulence will change direction on time
scales $\sim l_A/c$ due to magnetic field tangling and 
at the same time will experience pitch-angle scattering 
on time scale $\sim F(\mu)^{-1} l_A/c$.
It basically
implies an average effective mfp that is a fraction of $l_A$,
$\lambda_{mfp} = \psi l_A$, where $\psi < 1$; additional 
mechanisms of pitch-angle scattering from kinetic/plasma effects
will further reduce $\lambda_{mfp}$ (Sect. 3.3).

\noindent
It is thus convenient to re-write relevant quantities in terms of
$\beta_{pl}= 2 \Gamma^{-1} c_s^2/V_A^2$ ($\Gamma = 5/3$)
and the particles mfp in terms of the 
Alfv\'en scale, $l_A$ :

\begin{equation}
D_{pp} \simeq
3 \sqrt{5 \over 6}
{{c_s^2 }\over{c}} {{\sqrt{\beta_{pl}} }\over{L_o}}
M_t^3 \psi^{-3} p^2
\label{dpppsi}
\end{equation}

\noindent
and 

\begin{equation}
l_A = L_o M_A^{-3} = {{ (6/5)^{3/2} L_o}\over{( \sqrt{\beta_{pl}} M_t )^3 }}
\end{equation}

\noindent
Note that the condition $\Delta p \ll p$ is satisfied for:

\begin{equation}
\psi >  \sqrt{ {{( 3 c_s \sqrt{6/5} )}\over{( c \sqrt{\beta_{pl}} )}} }
\end{equation}

\subsection{A note on the role of kinetic effects}

The role of kinetic effects on small scales
is different for the turbulent
reconnection (see LV99, Kowal et al. 2009, Eyink et al. 2011, 2013, 
Eyink 2015, Lazarian et al. 2015) and tearing reconnection (see Loureiro
et al. 2007, Uzdensky et al. 2010, Ng et al. 2015).
Kinetic effects in the latter process accelerate the reconnection rate
inducing tearing of the current sheet. On the contrary, turbulent
reconnection is independent of the detailed kinetic physics, as follows
from theory (LV99, Eyink 2015) and confirmed in numerical
simulations where plasma physics effects were introduced using
anomalous resistivity (see Kowal et al. 2009).
Both approaches can be complementary with tearing destabilizing
the reconnection sheet if the level of turbulence in the system is
low. The outflow from the reconnection layer for extended current sheets
has the thickness $\Delta$ and the {\it Reynolds} number 
${\it Re} \sim V_A \Delta/\nu$, where $\nu$ is viscosity, which
for the ICM should be taken the perpendicular Braginsky viscosity.
With tearing inducing $V_{rec} > 0.01V_A$, which in turn is of the order 
of $V_A \Delta/l_x$ (eq.\ref{vrec}), $\Delta$ gets to be $> 0.01\, l_x$, 
making the {\it Reynolds} numbers of the outflow sufficiently large.
This should make the outflow turbulent and is likely to transfer the
reconnection into a regime similar to turbulent reconnection.

The physics of acceleration in turbulent and tearing reconnection in 3D
is very similar. The difference initially were due to 
the fact that acceleration in turbulent reconnection was considered
in 3D (de Gouveia dal Pino \& Lazarian 2005) while acceleration in
tearing was considered in 2D (Drake et al 2006).
As the actual 3D geometry of astrophysical reconnection is being
acknowledged and tearing reconnection layers demonstrate
more turbulence as the numerical resolution increases, strong
similarities emerge between the two mechanisms with the acceleration
being essentially related to the shrinking of 
3D loops\footnote{a comparison of particle
acceleration rates in 2D and 3D configurations is
discussed in Kowal et al. (2011)}.

\noindent
It is also true that particle acceleration and transport in reconnection 
regions are potentially sensitive to details of the physics of
reconnection (eg., Daughton et al 2011).
More recently the controversy on whether the tearing reconnection is
second or first order has emerged.
For example Drake et al (2013) find a shortening 
of magnetic island field line length because of adjacent islands
merging. This causes an increase in the parallel particle 
velocity and a decrease in perpendicular energy suggesting second order 
rather than first order Fermi mechanisms.
In fact attempts
to model second order Fermi acceleration of particles
diffusing across multiple reconnecting regions have been developed   
in the case of collisionless tearing reconnection mediated by
quasi 2D magnetic islands (eg. Le Roux et al 2015).

\noindent
In Sect.3.2 we have assumed first order acceleration in reconnection
regions because this mechanism has been clearly demonstrated 
in numerical simulations of turbulent reconnection (Kowal et al 2012).
We would believe that a similar effect should be present also
in tearing reconnection.
While the relative importance of the second and first order Fermi
acceleration in tearing reconnection may need to be settled in future,
for our treatment of the acceleration this is not so vital
and in principle
our picture can be extended also to the case of tearing reconnection. 
In fact, we only require that a particle leaving the reconnection region
gained the momentum $\Delta p$ whatever is the cause of the acceleration
\footnote{In the case of second order acceleration 
the term $dp/dt$ in our eq.\ref{dpdt} can be seen as 
the {\it systematic} part of a
second order process occurring within reconnection regions, i.e.
$dp/dt \sim {\tilde D}_{pp}/4p$ where ${\tilde D}_{pp}$ is the diffusion
coefficient in the particles' momentum space due to such a hypothetical
second-order mechanism on smaller scales}.
Once CRe leave reconnecting regions they will interact with 
the dynamics of magnetic field lines on very large scales, 
$l_A \sim$ 100 pc -- kpc.
On these large scales reconnection events are part of the turbulent cascade 
and we can use transport properties that follow from MHD turbulence
(Sect. 3.2).

The microphysics of the plasma may still be important for particle
acceleration and reacceleration even if the reconnection rates are 
determined by the LV99 mechanism.
In fact kinetic/plasma effects on small scales may induce 
pitch-angle scattering events in addition to those due to the
MHD fluctuations on large scales considered in Sect.3.2.
Overall the combined effects of scatterings on small and large
scales can be parameterized in the form 
of the particles mfp, $\lambda_{mfp}$, and in general scatterings on
small scales are expected to reduce $\lambda_{mfp}$ with respect to
that in Sect.3.2 and increase the acceleration rate.
We note that the small-scale physics is likely to be important 
when we consider the acceleration of particles from the thermal pool.
In this rescpect, potentially kinetic/plasma effects 
might provide also a solution to the problem of {\it seed}
particles in reacceleration models (Sect.1), however we do
not address the issue of the acceleration of energetic particles from
the thermal pool in this paper.

\section{Diffuse radio emission in galaxy clusters}

\subsection{Current view}

Turbulence and shocks are generally invoked to play a role
for the origin 
of radio halos and radio relics in galaxy clusters (BJ14 for review).
Nevertheless recent observations of the radio relic in A2256 unveiled a
complex/filamentary morphology of the radio emission and no
clear evidence for a direct connection with a shock, hinting a
possible role of magnetic 
reconnection (Owen et al 2014).
A role of magnetic reconnection for the origin of non-thermal
components in galaxy clusters was also envisaged on theoretical
grounds by Lazarian \& Brunetti (2011).

Radio halos are the most prominent non-thermal sources in galaxy clusters.
A popular scenario for their origin is based on the
hypothesis that turbulence generated during cluster mergers
(re)accelerates seeds electrons on Mpc scales 
(Brunetti et al 2001, Petrosian 2001, BJ14 for review).
Crucial ingredients of this scenario are however poorly
known. Primarily the challenge is to understand 
the chain of mechanisms that allow to drain energy from large scale
motions into mechanisms acting on smaller scales and that in turn
determines the efficiency of acceleration (see Brunetti 2016 for a recent
discussion).

\noindent
Presumably several turbulent components are generated in the ICM, both at
large and very small scales,
and all these components should jointly contribute to the
scattering and (re)acceleration of relativistic particles.
In the last years much attention has been devoted to 
the role of compressible turbulence that is driven at large scales
in the ICM from cluster mergers and that cascades at smaller
scales (Cassano \& Brunetti 2005, Brunetti \& Lazarian 2007, 2011a).
This is the simplest scenario that can be thought, 
nevertheless it predicts a 
straightforward connection between cluster mergers and radio halos
in agreement with observations. This is also 
the scenario 
adopted in follow-up numerical simulations that aim at investigating 
the origin of diffuse radio emission in
galaxy clusters (Donnert et al 2013, Beresnyak et al 2013, 
Donnert \& Brunetti 2014, Miniati 2015, Pinzke et al 2015).

On the other hand, recent numerical simulations of cluster 
formation have shown that most of the turbulence in galaxy clusters
is solenoidal (see Miniati 2015).
Although these simulations show that compressive 
turbulence is also generated in connection with clusters mergers,
this component is found fairly short living ($\leq$Gyr) and with a
kinetic spectrum that is steeper than that previously assumed 
in calculations of turbulent acceleration. The consequence is a 
reduction of
both the reacceleration period and the acceleration rate.
In reality these simulations do not tell us much about the microphysics
(eg. particles collision frequency) and power spectrum of electromagnetic 
fluctuations that both govern the acceleration efficiency and
{\it di per se} do not challenge acceleration models
based on compressive turbulence generated at large-scale (Miniati 2015,
Brunetti 2016).
However these are clearly thought-provoking results that motivate
us to investigate possible mechanisms that 
drain a fraction of the solenoidal turbulence in the ICM into particle
acceleration.

\begin{figure}
\begin{center}
\includegraphics[width=0.425\textwidth]{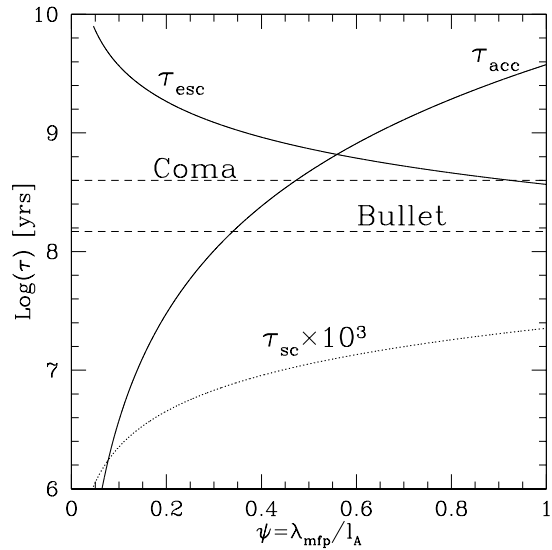}
\caption[]{
Acceleration time-scales as a function of $\psi$.
Horizontal dashed lines are the maximum reacceleration times,
$\tau_{max}$, 
requested for the radio halos in the Coma and Bullet clusters 
(Eq.\ref{tauconstrain}, see text).
Diffusion time on $L_{RH}=1$ Mpc (eq.\ref{escape}) and
scattering time-scale, $\lambda_{mfp}/c$ (multiplied by 1000 for
display proposal), are also reported.
Calculations are obtained assuming 
$M_t^2 =0.25$, $L_o = 300$ kpc, $\beta_{pl} = 60$, $c_s = 1500$ km/s.}
\label{Fig.Lr_RH}
\end{center}
\end{figure}

\subsection{A new mechanism ?}

In this respect the mechanism described in Sect.3 is particularly
attractive as it is based on super-Alfvenic solenoidal 
turbulence.
If we restrict to the situation of {\it fast} diffusion regime,
the acceleration time is (Eq.\ref{dpppsi}) :

\begin{equation}
\tau_{acc} = {{p^2}\over{4 D_{pp}}}
\simeq 
{{\sqrt{6/5}} \over {12}}
{c \over{c_s^2}} {{L_o}\over{\sqrt{\beta_{pl}}}}
M_t^{-3} \psi^3
\label{tauaccfin}
\end{equation}

\noindent
and the acceleration rate (or efficiency, throughout this
paper) $= \tau_{acc}^{-1}$.
The acceleration time 
is reported in Fig.1 as a function of the CRe mfp, $\psi=
\lambda_{mfp}/l_A$, assuming typical conditions in the ICM; the
acceleration efficiency rapidly increases with decreasing CRe mfp.
Fig.2 shows the effect induced on the acceleration time 
by increasing the turbulent Mach number and $\beta_{pl}$: 
for given values of $\psi$ and of the turbulent injection scale, 
$L_o$, the acceleration rate increases with both
Mach number and $\beta_{pl}$.
Calculations in Fig.2 are obtained
for typical conditions in the ICM, specifically we assume
a range of $\beta_{pl}\sim 30-120$ that is inferred using values
of magnetic field derived from Faraday rotation
measures (eg., Carilli \& Taylor 2002,
Feretti et al 2012) and turbulent injection
scales $L_o \geq 100$ kpc (BJ14, Br\"uggen \& Vazza 2015).

\noindent
In our model the mfp is a free parameter that is
difficult to calculate from first principles due to
our ignorance of kinetic effects on small scales 
(see Sect.3 for constraints based on MHD and 3.3 for a
discussion on kinetic effects).
However we can also infer independent constraints on that.
Statistics of radio halos suggests that they are
generated in massive mergers for a significant fraction of the
duration of these events (eg., Brunetti et al 2009,
Cassano et al 2013, Cuciti et al 2015).
It implies that CRe  
should be trapped within the emitting,
Mpc-scale, volume for $\geq$ few Gyr. This provides a limit to the diffusion
time of CRe on these scales and thus on their maximum mfp.
The time needed by CRe to 
diffuse on the scale of a radio halo, $L_{RH}$, is :

\begin{equation}
\tau_{esc} \sim {{L_{RH}^2}\over{4 D}}
\sim 
{5 \over 8} \sqrt{{ 5 \over 6}}
{1 \over c}
\left(
\sqrt{\beta_{pl}} M_t \right)^3
{{L_{RH}^2}\over{L_o}} \psi^{-1}
\label{escape}
\end{equation}

\noindent
that is reported in Fig.1.
The minimal condition $\tau_{esc} \geq$ few Gyrs implies
$\lambda_{mfp} \ll$ kpc, or in terms of MHD scale,
$\psi < 0.5 (\beta_{pl}/60)^{3 \over 2} (M_t/0.5)^3
(L_o/300 {\rm kpc})^{-1}$ (Fig.1, Eq.\ref{escape}),
i.e. larger values of the turbulent Mach number and of
the $\beta_{pl}$ constrain larger values of $\psi$. 
We note that the ratio of acceleration and
diffusion time, $\tau_{acc}/\tau_{esc} \propto \psi^4 L_o^2
M_t^{-6}$, rapidly decreases with increasing the turbulent Mach
number for a given value of $\psi$ and of the injection scale $L_o$
(Fig.2).

\noindent
Finally, 
we note that very small values of the mfp, that imply faster acceleration
(Eq.\ref{tauaccfin}),
are also ruled out from the condition $\Delta p \ll p$ (Eq.10).

\begin{figure}
\begin{center}
\includegraphics[width=0.425\textwidth]{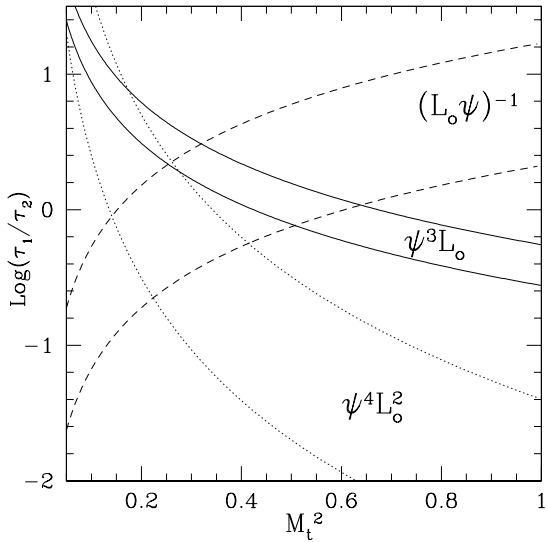}
\caption[]{
Ratio of model acceleration time and the $\tau_{max} =$150 Myrs  
for the Bullet radio halo (solid lines), and
ratio of model acceleration time and diffusion time on 1 Mpc (dotted lines) 
as a function of the turbulent Mach number.
Calculations assume a reference value $\psi =0.5$, and $\beta_{pl}=$30
and 120 (from top to bottom); other parameters being equal to Fig.1.
Dashed lines are the diffusion times in Gyr (from top to bottom with
decreasing $\beta_{pl}$).
Dependences on $\psi$ and $L_o$ are shown in the panels.}
\label{Fig.Lr_RH}
\end{center}
\end{figure}

Given these constraints on $\lambda_{mfp}$ and $\psi$,
the following point is to understand whether 
our mechanism can explain the diffuse radio emission observed
in galaxy clusters.
To fully 
address this point Fokker-Planck time-dependent calculations are necessary
(Eq.\ref{FP}).
However, as we anticipated in Sect.3.2,
numerous papers carry out Fokker-Planck calculations of particle 
reacceleration in galaxy clusters in the case $D_{pp} \propto p^2$.
These results can be automatically extended to our model and 
for this reason in this exploratory paper we do not focus on
Fokker-Planck analysis.

\noindent
In practice two main processes oppose to the reacceleration 
of CRe. 

\noindent
Synchrotron and inverse Compton losses balance the acceleration
of CRe channelling an increasing fraction of the energy gained by CRe
into radiation.
The time-scale of radiative losses (radiative life-time) of
CRe is :

\begin{equation}
\tau_{rad} (\gamma)= \gamma/{{ d \gamma}\over{d t}}
= {{6 \pi m_e c}\over{\sigma_T}} {{\gamma^{-1}}\over{B^2 +
B_{cmb}^2}}
\label{radiative}
\end{equation}

\noindent
where $B_{cmb} \simeq 3.25 (1+z)^2 \mu$G is the equivalent
field due to inverse Compton scattering with photons of the Cosmic
Microwave Background. In the ICM the CRe that produce synchrotron
radiation in the radio band have energies $\sim$ few GeV and
radiative life-times $\tau_{rad} \sim$ 100 Myrs (BJ14 for review).
The acceleration mechanisms that generate 
steep spectrum radio emission in galaxy clusters should have 
an efficiency that is comparable to that of the radiative losses of the
CRe that emit the observed radio emission.
Consequently, due to the balance between acceleration and losses,
after a reacceleration period of about 100 Myrs the spectrum of radio
emitting CRe evolves slowly with time and 
approaches quasi-stationary conditions, provided that
the acceleration rate is constant (BJ14 for review and references
therein).

\noindent
The second process that opposes to reacceleration
is more subtle and is due to physical mechanisms
that may damp the turbulent spectrum reducing with time the energy flux 
that becomes available for the reacceleration of CRs. In this respect 
the most obvious process is the damping (via collisionless dampings) 
of turbulence due to the same CRs that are reaccelerated.
This mechanism may affect (reduce) the acceleration rate on 
longer timescales (BJ14 for review and references therein).

\noindent
In this Section we consider only the first damping mechanism, due to 
radiative losses of CRe, whereas in
Sect.5.2 we will discuss the turbulence damping.

\noindent
In practice, the most relevant point here is to check whether our mechanism
allows reacceleration of CRe to the energies that are 
necessary to produce the synchrotron radiation observed in galaxy
clusters.
In Figs.1 and 2 the acceleration time is compared with
the acceleration time that is requested for radio halos.
This is not necessarily equivalent to the radiative cooling time of the
CRe emitting at the observed radio frequencies, because radio
halos have very steep synchrotron spectra, $\alpha \sim 1-2$ (flux
$\propto \nu^{-\alpha}$).
In this respect we use results from 
Cassano et al.(2010) that calculated synchrotron spectra 
in homogeneous models of compressive turbulent reacceleration 
(with $D_{pp} \propto p^2$) 
assuming reacceleration periods in the range $\Delta t/\tau_{acc} \sim 2-4$. 
They found that the spectra become steeper
than those of typical radio halos at frequencies greater than   
$\nu_s \sim \xi \nu_b$, where $\xi \sim 6-8$,
and $\nu_b$ is the critical synchrotron frequency emitted by 
electrons having acceleration time equal to the
time-scale of radiative losses, i.e. CRe with Lorentz factor 
$\gamma_b = {{6 \pi m_e c}\over{\sigma_T}} \tau_{acc}^{-1}/(B^2 + 
B_{cmb}^2)$.
The highest frequency is generated in a magnetic field $B \sim 
B_{cmb}/\sqrt{3}$, since this is the field where
CRe emitting at a given frequency have maximum lifetime.
It leads to a {\it minimal} request for a mechanism being sufficient
to reaccelerate CRe emitting at the observed frequency
$\nu_o = \nu_s/ (1+z)$.
This condition
corresponds to a {\it maximum} reacceleration time, $\tau_{max}$ :

\begin{equation}
\tau_{acc} \leq \tau_{max} = 156 \left(
{{\xi }\over{\nu_o(GHz)}} \right)^{1 \over 2}
(1+z)^{- 7/2} \,\,\,\,\, {\rm (Myr)}
\label{tauconstrain}
\end{equation}

\noindent
In Figs.1 and 2 we report relevant values of $\tau_{max}$ derived from
Eq.\ref{tauconstrain}. These values are
derived for the radio halos in the Coma and Bullet clusters,
$\tau_{max} = 400$ and $= 150$ Myr, respectively,
by considering observed steepening frequencies
$\nu_o \simeq 1$ GHz for Coma
(eg. Brunetti et al 2013 and references therein)
and $\simeq 5$ GHz for the Bullet cluster
(a hint of steepening is observed in this cluster at $\sim 8$ GHz,
Liang et al 2000, Shimwell et al 2014, Marchegiani \& Colafrancesco
2015).

\noindent
The comparison between $\tau_{max}$ and the reacceleration time
predicted by our model selects the parameters space ($M_t$, $\psi$,
$L_o$) that allows to explain radio halos.
Specifically, 
the acceleration time predicted by our model (Eq.\ref{tauaccfin})
leads to a steepening frequency :

\begin{eqnarray}
\nu_s ({\rm GHz})
\simeq 2.45  \left(
{{c_s}\over{10^3 {{ {\rm km} }\over{ {\rm s} }} }}
\right)^4
{{\beta_{pl}}\over{10^2}}
{{\xi }\over 7}
\left(
{{10^2 {\rm kpc}}\over{L_o}}
\right)^2
\left(
{{ M_t}\over{0.5}}
{{0.5}\over{\psi}}
\right)^6 \nonumber \\
\times
\left( 
{{ B_{\mu G} }\over{(B^2_{\mu G} + B^2_{cmb})^2}}
\right)/0.01
\label{nusgeneral}
\end{eqnarray}

\noindent
implying that an emitting synchrotron frequency 10 times larger is 
produced by increasing the turbulent Mach number (or reducing the CRe
mfp) by about 50 percent, or by assuming a turbulent-injection scale
that is 3 times smaller.
For $B \simeq B_{cmb}/\sqrt{3}$, the 
Mach number and the observed steepening frequency in our model
are connected via (from Eq.\ref{nusgeneral}) :

\begin{equation}
M_t \simeq 0.8
{{10^3 {{ {\rm km} }\over{ {\rm s} }} }\over {c_s}}
\left(
{{\nu_o({\rm GHz})}\over{35}}
{ 7 \over \xi}
\right)^{1 \over 6}
\left(
{{L_o}\over{10^2 {\rm kpc}}}
\right)^{1 \over 3}
{{
(1+z)^{7 \over 6}
{{\psi}\over{0.5}} }\over
{( {{n_{ICM} }\over {10^{-3}}})^{1 \over 6}}}
\label{mach}
\end{equation}

\noindent
that allows us to conclude that turbulent Mach
numbers $M_t^2 \sim
0.2-0.5 (\psi/0.5)^2 (L_o/ 300 \,{\rm kpc})^{2/3}$ are sufficient 
to reaccelerate CRe that generate synchrotron radio emission 
in hot, ${\rm kT} \sim 5-15$ keV, clusters. 
Note that under these conditions the model satisfies also 
the additional requirements that are necessary 
to explain radio halos and cluster-scale 
radio emission, specifically the acceleration time is shorter than the 
diffusion time of CRe ($\tau_{acc} < \tau_{esc}$) and 
CRe are efficiently trapped within Mpc-scale emitting regions, 
i.e. $\tau_{esc} >$ few Gyrs (Figs.1 and 2).
We note that the above amount of turbulent energy/Mach numbers is
consistent with that found in high-resolution
cosmological (fluid) simulations (eg., Vazza et al 2011, Miniati 2015).

\noindent
Interestingly the same turbulent motions that are used in our
model to reaccelerate particles will induce a broadening of the
X-ray lines (for example the FeXXV lines at 6.7 keV) that can be tested
by the forthcoming ASTRO-H satellite (Kitayama et al 2014) and, in the not too
distant future, by ATHENA (Ettori et al. 2013).
In particular, assuming a typical electron number density
$=10^{-3}$cm$^{-3}$, from Eq.\ref{mach} we get {\it minimum} turbulent
velocities at the scale $L_o$, $\delta V_o \sim 700$ and 
$450 (\psi/0.5) (L_o/100 {\rm kpc})^{1/3}$ km/s for the halo in the Bullet and 
Coma clusters, respectively
(corresponding to Mach numbers $\simeq 0.35$ and 0.3, respectively)
(see also Sect. 5).

At this point it is worth to comment that
our conclusions are based on results from compressive 
turbulent acceleration models in the homogeneous case and on their scaling
to our model. 
Follow up Fokker-Planck time-dependent
calculations combined with numerical simulations 
are required to account for non-homogeneous distributions of the physical 
parameters in the turbulent ICM (see discussion in Sect. 4.3) and to 
model the time evolution of the synchrotron spectrum of cluster-scale 
emission. For example, with regard to this latter point, 
the parameter $\xi \sim 6-8$ in Eqs.\ref{tauconstrain}--\ref{mach} is 
derived for acceleration periods
$\Delta t/\tau_{acc} \sim 2-4$ (Cassano et al 2010) (i.e. $\Delta t
\sim 0.3 - 1$ Gyr for typical values of $\tau_{acc}$), corresponding 
to a situation where radio spectra evolve slowly with time
(quasi-stationary)
because reacceleration is balanced by the energy losses of the CRe emitting
in the radio band.
On the other hand the shape of the spectrum of radio halos may be
different during earlier and/or later phases.
During earlier phases reacceleration is not balanced yet by losses
and the spectrum rapidly evolves with time being very steep (implying 
smaller $\xi$) at the beginning of the
reacceleration phase ($\Delta t/\tau_{acc} < 2$).
Similarly, at later stages the evolution of CRe spectrum 
is potentially affected by saturation effects due to turbulence
damping (see Sect. 5).

\subsection{A note on the spectrum of radio halos}

The spectral shape 
of radio halos provides information that go beyond 
simple arguments based on acceleration time-scales.
Unfortunately due to observational limitations
a spectrum with at least three
measurements is available only for a few halos, and 
significant scatter of measurements is seen 
in the cases where 
the data-points cover at least 1 order of magnitude in frequency range
(e.g. Macario et al 2013, Shimwell et al 2014).
Current studies suggest that 
the spectra of radio halos are quite different.
Assuming a simple power-law shape (in the form 
flux $\propto \nu^{-\alpha}$) the 
spectral slopes measured between two typical
frequencies (0.3 -- 1.4 GHz) are currently found 
in the range $\alpha \sim 1-2$ (eg Venturi 2011). 
The Coma radio halo is a unique case with several 
spectral measurements covering about 2 orders of magnitude in 
frequency. Despite the scatter that is observed between the different
measurements a spectral steepening is observed at higher frequencies, 
suggesting a corresponding break in the spectrum of the emitting
electrons (Brunetti et al 2013 and references therein). 

We still lack a comprehensive understanding of the spectrum of radio halos
predicted in reacceleration models. 
In the context of our model and in the other turbulent reacceleration models,
the spectrum of radio halos results from the interplay between
the CRe spectrum and the probability distribution function (PDF) of
magnetic field strengths in the emitting volume.
Simple homogeneous reacceleration models, where both 
the magnetic field and acceleration rate are constant
in the emitting volume, predict a curved
spectrum due to the interplay between energy losses and
diffusion in the particle's momentum space (e.g. Cassano \& Brunetti
2005).
More realistic turbulent
models account (at least) for the decline of the magnetic field
strength with radius in clusters.
The effect is a stretching in frequency of the volume-integrated
synchrotron spectrum of the halos that allows to successfully explain 
observed spectra (BL07, Brunetti et al 2008,
Brunetti \& Lazarian 2011a, Macario et al 2013, Donnert et al 2013).
In the real case of a turbulent ICM the magnetic PDF is expected to be
broad, and the turbulent and plasma-collisional properties can vary in
the emitting, Mpc, volume. This, combined with the possibility
of CRe diffusion across different turbulent regions, is expected to 
stretch considerably the synchrotron spectrum in frequency.
Examples of synchrotron spectra from non-homogeneous turbulent
reacceleration can be found in ZuHone et al (2013), where
turbulent acceleration with $D_{pp}\propto p^2$ and synchrotron emission
from mini-halos (cluster's core) are calculated using high-resolution 
(kpc) MHD simulations combined with tracer particles to follow the
dynamics and evolution of CRe.

\section{Discussion on CRs reacceleration in the ICM}

\subsection{A comparison with reacceleration by 
compressive turbulence}

In this Section we compare our model with 
stochastic reacceleration models by compressible turbulence in the ICM.

We start from the 
the momentum-diffusion coefficient that is expected 
in the classical Fermi theory. 
This is (e.g., Longair 2011):

\begin{equation}
D_{pp} \sim 
p^2 {{\delta V_l^2}\over{ c l }}
\sim 
p^2 {{V_{ph}^2}\over{c l}} 
{{(\delta B)_l^2}\over{B_0^2}} 
\label{fermi}
\end{equation}

\noindent
where $l$ is the minimum scale of eddies,
$V_{ph}$ is the phase velocity of waves and in high-beta plasma
conditions it is $\delta B/B_0 \sim \delta V/V_{ph}$.
It follows that 
$D_{pp}/p^2$ is simply $c/\delta V_l$ times the turn-over time 
of turbulent eddies at the dissipation scale.

These dependences basically apply also to the case of the 
interaction between compressible turbulence (essentially
fast modes) and particles via Transit-Time-Damping (TTD).
TTD interacts through the $n=0$ Landau resonance (eg. Fisk 1976,
Schlickeiser \& Miller 1998).
This interaction is essentially the coupling between the magnetic 
moment of particles and the parallel magnetic field gradients and
it uses the turbulent 
energy that is available on a broad range of scales, in principle
between the dissipation and the injection scales.
The diffusion coefficient in the particles momentum space due to TTD
in the ICM is derived in BL07 :

\begin{eqnarray}
D_{pp} =
{{\pi^2}\over{2c}}
{{p^2 c_s^2}\over{
B_0^2}} \int_{c_s/c}^1
d\mu {{1 - \mu^2}\over{\mu}}
\left(
1 -
({{c_s}\over{c \mu}})^2
\right)^2 \times
\nonumber\\
\int_{k_0}^{k_{cut}}
dk W_B(k) k
\label{dppTTD}
\end{eqnarray}

\noindent
Similarly to the classical second order Fermi mechanism, 
the acceleration depends on  
the minimum/dissipation scale of the electromagnetic
fluctuations and on the energy density of these fluctuations, being  
$\int dk W_B k \sim k_{cut} \delta B^2_{k_{cut}}/8\pi$.

\noindent
The dissipation scale is set by the competition between
turbulent cascade and collisionless damping. Two situations are 
considered in the literature:

\begin{itemize}

\item[{\bf (i)}] the collision frequency of thermal
particles is $\omega_{ii} < \omega=k c_s$. In this case
the damping of the waves is dominated by TTD with thermal particles
(electrons) and $k_{cut} \sim 10^4 k_o M_t^4$ (BL07,
Miniati 2015, Brunetti 2016). We stress that under these conditions
{\it most} of the turbulent energy is dissipated into the heating
of the thermal ICM (see BL07).
Assuming a Kraichnan spectrum, this leads to a reacceleration time for
CRe :

\begin{equation}
\tau_{acc} = {{p^2}\over{4 D_{pp}}} 
\simeq 125 
({{M_t}\over{1/2}})^{-4} \left(
{{L_o/300 \, {\rm kpc}}\over{c_s/1500\, {\rm km \,s^{-1}}}}
\right)
\,\,\,\, ({\rm Myr})
\label{timeaccC}
\end{equation}

\item[{\bf (ii)}] the collisions between thermal 
particles in the ICM are fast, $\omega_{ii} > \omega=k c_s$. 
This is the case where 
scatterings are mediated by magnetic field perturbations
driven by plasma instabilities (eg., Lazarian \&
Beresnyak 2006, Brunetti \& Lazarian 2011b,
Santos-Lima et al 2014 and references therein). 
Contrary to case (i), here 
the damping of the waves is due to TTD with relativistic
particles (CR protons and CRe), 
turbulence reaches smaller scales and the reacceleration 
time (for Kraichnan spectrum) is (Brunetti \& Lazarian 2011b) :

\begin{equation}
\tau_{acc} \simeq 6 
({{M_t}\over{1/2}})^{-4} \left(
{{L_o/300 \, {\rm kpc}}\over{c_s/1500\, {\rm km \,s^{-1}}}}
\right)
\left( {{{\cal R}_{CR} }\over{25}} \right)
\,\,\,\,\,\, ({\rm Myr})
\label{timeaccCR}
\end{equation}

\noindent
where 

\begin{equation}
{\cal R}_{CR}
= \sum_{i=e,p}
{{c \int p^4 dp {{\partial f_{i}(p)}\over{\partial p}} }\over
{\rho_{ICM} c_s^2 }} 
\end{equation}

\noindent
$\rho_{ICM}$ is the thermal density of the ICM, 
and $f_i(p)$ is the distribution function of CRs (CRe and CR 
protons) in the momentum space.

\end{itemize}

\noindent
The two acceleration schemes have been used estensively to
explain radio halos and mini radio halos in galaxy clusters 
(BL07, Brunetti \& Lazarian 2011a,b, Beresnyak et al 2013,
ZuHone et al 2013, Donnert et al 2013, Donnert \& Brunetti 2014,
Miniati 2015, Pinzke et al. 2015).
The main assumptions in these studies is that compressible turbulence 
is a relevant part of turbulence in the ICM.
Challenges of these models due to our poor understanding of the ICM
microphysics have been extensively discussed in Brunetti
(2016). In particular, it has been shown that an increasingly steeper 
spectrum of the electromagnetic fluctuations induces a slower
acceleration rate essentially because fluctuations are dissipated at
larger scales. 
Recent cosmological simulations suggest that 
compressive turbulence in the ICM is 
Burgers-like ($W \propto k^{-2}$). In this case (assuming that
velocities and electromagnetic fluctuations follow the same
scalings), for typical velocities
of the turbulent eddies $\delta V \sim 500-800$ km s$^{-1}$ on
large scales ($L_0 =300$ kpc), the efficiency
of TTD acceleration 
drops by about 10 times with respect to that predicted adopting a Kraichnan
spectrum (Brunetti 2016, their Fig.1). 
Under these unfavourable 
conditions in the collisionless scenario (case (i), Eq.\ref{timeaccC}) 
TTD would become inefficient and would not provide a valuable scenario 
for the origin of cluster-scale radio emission. In this case
explaining radio halos would 
require that the ICM behaves mostly collisional (i.e. case (ii),
Eq.\ref{timeaccCR}) (see Miniati 2015).

As already mentioned the main motivation for the present paper is that
large-scale turbulence in the ICM is mainly solenoidal (Miniati 2014). 
This incompressible turbulence is dissipated into the thermal 
ICM at a much smaller rate than
the compressible part and consequently it
provides a previously unexplored, natural energy reservoir for stochastic 
particle acceleration in the ICM.
In this paper we essentially suggest that part of this energy drains
into relativistic particles via the interaction with magnetic
field lines diffusing in turbulent reconnection.
Note that 
the acceleration mechanism explained in Sect.3.2 do not drain turbulent
energy into thermal ICM as these particles do not diffuse on
sufficiently large scales.
Our mechanism also differs from other models for diffuse radio emission from 
galaxy clusters that are based
on particles-waves interaction via gyro-resonance, where  
the non-linear coupling between particles and Alfv\'en waves
at very small, resonant, scales $k \sim eB E^{-1}/(\mu \pm V_A/c)$ 
is used (Ohno et al 2002, Fujita et al 2003, Brunetti et al 2004,
Fujita et al 2015).
In fact it is unlikely that gyro-resonant acceleration is driven
directly by the cascade of incompressible turbulence from large-scales
to very small scales because the scale-dependent
anisotropies that are developed in the Alfv\'enic cascade at MHD scales
quench the acceleration (Chandran 2000, Yan \& Lazarian 2004).

\noindent
In our model the 
acceleration drains energy essentially from the
Alfv\'enic motions at the
Alfv\'en scale, $l_A$. 
In principle this energy reservoir is large and of the
order of the energy flux associated to the turbulence at larger scales,
$V_A^3/l_A \sim \delta V^3/L_0$ for Kolmogorov spectrum.
As a reference, by considering
$\psi \sim 0.2-0.5$ and typical conditions for the solenoidal
turbulence in the ICM, $M_t^2 \sim 0.3-0.8$ 
and $L_o = 100-500$ kpc (consistent 
with numerical simulations), the acceleration rates
(Sect. 4.2) are similar to those
expected in the case of collisionless TTD with compressive turbulence
for Kraichnan spectrum (case (i), Eq. \ref{timeaccC})\footnote{this
is also equivalent to acceleration rates 
expected in the case of collisional TTD ((ii), eq.\ref{timeaccCR})
assuming Burgers--like spectrum of turbulence, or
in the case of Kraichnan spectrum assuming smaller values of the 
turbulent Mach number, $M_t^2 \sim 0.05-0.1$}.
Since collisionless TTD has been successfully applied to radio halos,
we conclude that our mechanism may provide a valuable 
alternative to the mechanisms based on compressible turbulence or that 
it may play a role in combination with these other mechanisms.

\subsection{Dampings and saturation effects}

TTD reacceleration by compressible turbulence in the ICM has been calculated 
considering self-consistently acceleration and turbulent-damping rates
(BL07, Brunetti \& Lazarian 2011a,b).
On the other hand in Sect.3 we have calculated the diffusion coefficient 
in the particles' momentum space in the test particle regime.
If a large fraction of the turbulent energy flux in our model is 
channelled into CRs, saturation effects become important.
Acceleration and turbulent damping, $\Gamma$, 
are connected through detailed balancing :

\begin{equation}
\sum_{i=e,p}
\int d^3p E {{\partial f_i}\over{\partial t}}=
\int d^3k W({\vec k}) \Gamma({\vec k})
\label{detailedbalance}
\end{equation}

\noindent
where the left-hand side of Eq.\ref{detailedbalance} accounts for the
energy channelled into CRs and the right-hand side accounts for
the energy that is extracted from turbulence via damping; $W({\vec k})$
is the 3D turbulent spectrum ($W({\vec k}) = 4 \pi k^2 W(k)$ in the
isotropic case).
Assuming pitch-angle isotropy of CRs, the time-evolution of the
CRs momentum distribution in Eq.\ref{detailedbalance} depends on the 
momentum-diffusion coefficient via :

\begin{equation}
{{\partial f_i}\over{\partial t}} =
{1 \over{p^2}} {{\partial }\over{\partial p}}
\left(
p^2 D_{pp} {{\partial f}\over{\partial p}}
\right)
\label{isotropy}
\end{equation}

\noindent
and the evolution of the turbulent spectrum is given by (eg. BL07 and
references therein) :

\begin{equation}
{{\partial W}\over{\partial t}} = {{\partial }\over{\partial k}}
\left(
k^2 D_{kk} {{\partial }\over{\partial k}} ( {{W(k)}\over{k^2}} )
\right) - \Gamma W(k) + I(k)
\label{turbulence}
\end{equation}

\noindent
where the first and second terms in the right-hand side account
for turbulent cascade and damping, respectively,
$D_{kk} \sim k^2/\tau_{kk}$ is the diffusion coefficient in the
wavenumber space\footnote{more specifically
it is $\tau_{kk} = k^3/ {{\partial }\over{\partial k}} (D_{kk} k^2)$},
$\tau_{kk}$ is the turbulent cascading-time at scale $k^{-1}$, and
$I(k)$ accounts for the process of
injection of turbulence at large scales.

\noindent
Eqs.\ref{detailedbalance}--\ref{turbulence} connect acceleration rate, CR
energy densities and turbulent damping and cascading. 
They essentially imply that 
the increase of the energy density 
of CRs with time produces a stronger damping, with the consequence that 
the turbulent spectrum is gradually suppressed and that 
acceleration is reduced.
In this regime full Fokker-Planck time-dependent 
calculations, combined with the modeling
of the evolution of the turbulent spectrum, are necessary for a correct
evaluation of the acceleration process (see e.g. Brunetti et al 2004,
Brunetti \& Lazarian 2011b for examples of
calculations in the ICM).
These saturation effects also limit 
the applicability of the approach used in Sect.4.2, that is based
on the scaling of the (spectral) analysis in compressive turbulent
acceleration to the case of (incompressible) turbulent reconnection.

\noindent 
In our acceleration model the 
turbulent eddies on scale $\sim l_A$ are the most 
important ones.  Consequently 
the importance of saturation effects can be evaluated by 
comparing damping and turbulent cascade at these scales:
if damping rate is smaller than the rate of 
turbulent cascade, $\tau_{l_A l_A}^{-1} \sim V_A l_A^{-1}$,
the turbulent spectrum is not modified and saturation effects are not
important. This is the 
case when the energy flux of turbulent cascade at the Alfv\'en scale,
$\sim \rho_{ICM} V_A^3 l_A^{-1}$, is much larger than 
the energy flux that is channeled into CRs.
The two quantities can be connected via :

\begin{equation}
\rho_{ICM} V_A^3 l_A^{-1} \eta_{CR} \sim 
c \int d^3p {1 \over p} {{\partial }\over{\partial p}}
\left(
p^2 D_{pp} {{\partial f}\over{\partial p}}
\right)
\label{balance}
\end{equation}

\noindent
where $\eta_{CR}$ is the fraction of turbulent energy flux that 
is transferred into CRs by our mechanism, and $D_{pp}$ is given in
Eq.\ref{dpppsi}.
The limit $\eta_{CR} \rightarrow 1$ corresponds to strong damping.
In fact this limit constrains the minimum value of
$\psi$ that is allowed in test-particle calculations 
(i.e. neglecting saturation/damping effects) of our model.
From Eqs.\ref{dpppsi} and \ref{balance} this is :

\begin{equation}
\psi_{min} \sim \left(
{5 \over 2} \beta_{pl}
{{V_A}\over{c}}
{{\epsilon_{CR} {\cal S}_{CR}}\over{
\rho_{ICM} c_s^2}}
\right)^{1/3}
\label{condition}
\end{equation}

\noindent
where we define :

\begin{equation}
{\cal S}_{CR} =
{{ \sum_{i=e,p}
\int d^3p {1 \over p} {{\partial }\over{\partial p}} (p^4
{{\partial f_i}\over{\partial p}} ) }\over{
\sum_{i=e,p} \int d^3p f_i(p) p}}
\label{factorf}
\end{equation}

\noindent
As expected, limits on $\psi$ depend on the energy density of CRs in
the ICM. Theoretical reasons suggest that CR protons are the most
important CR component in galaxy clusters (BJ14 for review).
In the recent years gamma-ray
and radio observations allowed to 
constrain the energy density that is associated to CR protons 
in the central Mpc-regions of galaxy clusters, $\epsilon_{CR}
<$ few $0.01 \times \epsilon_{ICM}$ (eg., Ackermann et al 2010, 
2014, Zandanel \& Ando 2014,
Brunetti et al 2007, 2008, BJ14 for review).
If we use these reference values, Eq.\ref{condition} basically implies 
$\psi > 0.1$.\footnote{On the other hand, if we consider only CRe 
in the ICM, or an {\it ad hoc} situation where CR protons are not
accelerated by our mechanism, it is 
${\cal S}_{CRe} \epsilon_{CRe} \sim 10^{-5}-10^{-4} \rho_{ICM} c_s^2$ 
and essentially $\psi > 0.02$.}
This limit on $\psi$ is fully consistent with the 
values used/derived in Sect.4.2 to explain
cluster-scale radio emission.

Interestingly for $\psi \sim 0.1$ the turbulent velocities that are
necessary to explain radio halos are still significant $\delta V_o
> 100$ km/s (for example $> 90$ and $> 140 (L_o/100)^{1/3}
(n_{ICM}/10^{-3})^{-1/6}$ km/s for the radio halos in Coma and
Bullet clusters, respectively, see Sect.4). 
This essentially sets the level of the minimum turbulent broadening of the
profile of X-ray lines that is expected in our model.
Non detection of such a broadening 
with future X-ray
spectrometers, for example with ATHENA, will rule out our model
under the conditions explored in this paper (namely fast diffusion
approximation and $\Delta p \ll p$, Sect.3.2)\footnote{alternatively it
will require
an {\it ad hoc} situation where CRp are not accelerated efficiently
by our mechanism}.

\subsection{Reacceleration of CR protons}

As a final remark we point out that 
in principle our mechanism reaccelerate also CR protons that will
accumulate their energy more efficiently than CRe (because 
CRp are subject to less efficient energy losses, see BJ14 for a review).
The energy density accumulated by CR protons in a reacceleration
time $\Delta t$ will be :

\begin{equation}
\rho_{ICM} V_A^3 l_A^{-1} {\langle \eta_{CR} \rangle}_{_{\Delta t}} 
\Delta t
\approx 
\Delta t \int d^3p E {{\partial f_p}\over{
\partial t}} \sim \delta \epsilon_{CR}
\label{Eaccumulated}
\end{equation}

\noindent
where ${\langle \eta_{CR} \rangle}_{_{\Delta t}} \Delta t =
\int_0^{\Delta t} \eta_{CR}(t) dt$.

\noindent
In practice CR protons will release part of this energy due to losses in
Gyr-time scales.
In a few Gyrs CR protons will be also advected and transported by large-scale
motions on scales larger than the relevant turbulent scales 
and thus energy will be redistributed on large-scales decreasing
$\delta \epsilon_{CR}$.
Nevertheless we can use Eq.\ref{Eaccumulated} under the condition
$\delta \epsilon_{CR} < \epsilon_{CR}$ to derive a (order of magnitude)
limit on $\Delta t$; it is useful also to check if our calculations
in the case of radio halos are self-consistent.
This is (from Eqs.\ref{balance}, \ref{dpppsi} and \ref{Eaccumulated}) :

\begin{equation}
\Delta t \sim {c \over 3} {{l_A}\over{V_A^2}} {{ \epsilon_{CR}
(\delta \epsilon_{CR}/\epsilon_{CR}) }\over{
{\langle \epsilon_{CR}(t) {\cal S} \rangle}_{_{\Delta t}} }} \psi^3
\label{deltat}
\end{equation}

\noindent
where $l_A = M_t^{-3} L_o (V_A/c_s)^3$. At this point Eq.\ref{deltat}
can be combined with the Mach number that is requested to generate 
radio halos with observed synchrotron steepening
frequency $\nu_o$ (under the minimal condition $B \sim
B_{cmb}/\sqrt{3}$, Eq. \ref{mach}) :

\begin{equation}
\Delta t \sim \left(
{{\nu_o(GHz)}\over{3}} {{ 7}\over{\xi}} \right)^{- {1 \over 2}}
(1+z)^{- {3 \over 2}}
{{ \epsilon_{CR}
(\delta \epsilon_{CR}/\epsilon_{CR}) }\over{
\langle \epsilon_{CR}(t) {\cal S} \rangle_{\Delta t} }}
\,\,\,({\rm Gyr})
\label{deltatfin}
\end{equation}

\noindent
where typically $\epsilon_{CR}/
{\langle \epsilon_{CR}(t) {\cal S} \rangle}_{_{\Delta t}}
\,\,\,\,\, \gtsim \,\,\, 1$. 
We thus conclude that radio halos observed at GHz frequencies 
can be generated by our mechanism for Gyrs. 
These time-scales are comparable to the
life-times of radio halos evaluated from statistical analysis (eg.
Brunetti et al 2009, BJ14) suggesting
that the exploratory (test-particle) 
calculations in Sect.3.2 and 4 provide a good approximation.
A reliable modeling of the acceleration process for 
longer acceleration periods (or for radio halos with spectra 
extending to $\nu_o \gg$ few GHz) requires
full Fokker-Planck time-dependent 
calculations combined with numerical simulations
of clusters (including CRs transport).

\section{Summary}

We suggest that relativistic particles diffusing 
in super-Alfvenic incompressible turbulence gain energy
due to the statistical interaction with field
lines in regions of magnetic reconnection and turbulent dynamo.
In view of large-scale turbulence being present in the ICM
we assume a scheme based on turbulent reconnection and assume
that MHD turbulence provides a good guide as far as reconnection
is considered.
We calculate the acceleration rate in the {\it fast} diffusion
regime and in the limit $\Delta p \ll p$.
Under these conditions the mechanism is essentially 
a second-order Fermi.

\noindent
In this exploratory paper we account for the interaction 
between magnetic field lines and CRe on large scales, $\sim$0.1-1 kpc,
assuming transport properties and turbulent scalings that follow
from MHD turbulence. Additional kinetic effects on smaller-scales
may affect the physics of reconnection and CRe transport and
acceleration. For this reason we consider CRe mfp as a free parameter
that is however bounded by the combination of basic constraints, including
the confinement of CRe (Sect.3.2), the 
interaction with large-scale fields (Sect. 3.2) and indirect
considerations
on the turbulent energy flux that is available (Sect.5).

\noindent
We propose that the mechanism may play a role for the
origin of radio halos and large-scale diffuse emission
in galaxy clusters. 
In fact, 
assuming reliable conditions in the ICM we have shown that the expected
reacceleration rate is similar to that of classical reacceleration
models proposed for radio halos, provided that the CRe mfp is a
fraction $\sim 0.1-0.5$ of the MHD scale. 
Smaller values of the mfp would make the mechanism even
more efficient, however in this case saturation/damping effects
(Sect. 5.2) should become important invalidating the test-particle
approach adopted in our paper and deserve future studies.

\noindent
Similarly to previous reacceleration 
models the diffusion coefficient in the momentum
space is $D_{pp} \propto p^2$ and thus, in principle 
all previous results based
on solutions of Fokker-Planck equations can be easily
extended to our case.
However our approach differs from these previous models 
because it is based on the incompressible part of the turbulence 
and because it is mediated by turbulent reconnection and dynamo.
According to current numerical cosmological simulations
incompressible turbulence in the ICM is found to be dominant,
super-Alfvenic and {\it quasi}--sonic and thus the new mechanism
explored in our paper appears particularly appealing.
Interestingly in our mechanism the conditions that are necessary 
to generate radio halos have unavoidable consequences on the
broadening of the profile of X-ray lines, allowing a test
of model assumptions with forthcoming X-ray calorimeters (ASTRO-H,
ATHENA).

\noindent
This is an exploratory paper.

\noindent
One of the critical points is the particles mfp.
The combined effect of super-Alfvenic motions and
resonant mirroring with pseudo Alfven modes constrains
the mfp $\sim$ a fraction of the MHD scale.
However in a more general situation additional scattering agents 
induced by additional turbulent
components may reduce the mfp; this might have the 
potential to increase the acceleration rate.

\noindent
Another critical point is the assumption of turbulent reconnection
scheme. The basic assumption here is that reconnection is part of
the MHD cascade.
This appears quite natural in the ICM given
the presence of large-scale turbulence that can be driven at large
scales in merging clusters and given
the observed connection between
mergers and non-thermal emission from galaxy clusters.
Other schemes, such as tearing reconnection, do not invalidate
{\it per se} our reacceleration scheme, provided that reconnection
and dynamo are still triggered by large-scale motions.
This however requires an extension of our modelling.

\noindent
Calculations in this paper focus on the hypothesis of fast-diffusion
regime and $\Delta p \ll p$ (Sect. 3.2).
Numerical simulations can be used to extend this exploratory study
to the situation $\Delta p \sim p$, in which case the mechanism is
expected to differ significantly from a second order Fermi.
Simulations will also allow to calculate the acceleration rate 
under more realistic configurations of reconnection and dynamo regions.

\noindent
Finally it is worth to mention that in our model we focus
on the reacceleration of CRe.
Also CR protons, if present, will be reaccelerated by our mechanism.
As a sanity-check we have shown that,
assuming the physical parameters that are requested to explain radio
halos, the energy budget of CR protons resulting from a reacceleration
period comparable to the life-time of halos is consistent with current
limits from $\gamma$-ray observations.
At this point, since reacceleration rates and diffusion coefficient, 
$D_{pp}\propto p^2$, in our model 
are equivalent/similar to those in lepto-hadronic reacceleration models
where CR protons and their secondary particles have been explicitly 
taken into account (eg. Brunetti \& Lazarian 2011a, Brunetti et al
2012, Pinzke et al 2015), also the conclusions of these
previous works can be extended to our picture.

\section{Acknowledgments}
We thank the referee for useful
comments that have improved the presentation of the paper.
GB and AL acknowledge support from the Alexander von Humboldt Foundation
and discussions with Prof. R. Schlickeiser. GB acknowledges support
from PRIN-INAF 2014, AL acknowledges support from NASA grant
NNX14AJ53G.


\begin{thebibliography}{}
\bibitem{} Ackermann, M., Ajello, M., Allafort, A., et al.\ 2010, \apjl, 717, L71 
\bibitem{} Ackermann, M., Ajello, M., Albert, A., et al.\ 2014, \apj, 787, 18 
\bibitem{} Beresnyak, A.\ 2013, arXiv:1301.7424 
\bibitem{} Beresnyak, A., Yan, H., \& Lazarian, A.\ 2011, \apj, 728, 60
\bibitem{} Beresnyak, A., Xu, H., Li, H., \& Schlickeiser, R.\ 2013, \apj, 771, 131
\bibitem{} Bhattacharjee, A.\ 2004, \araa, 42, 365 
\bibitem{} Br{\"u}ggen, M., \& Vazza, F.\ 2015, ASSL, 407, 599
\bibitem{} Birn, J., Drake, J.~F., Shay, M.~A., et al.\ 2001, \jgr, 106, 3715 
\bibitem{} Brunetti, G.\ 2016, Plasma Physics and Controlled Fusion, 58, 014011 
\bibitem{} Brunetti, G., Setti, G., Feretti, L., \& Giovannini, G.\ 2001, \mnras, 320, 365
\bibitem{} Brunetti, G., Blasi, P., Cassano, R., \& Gabici, S.\ 2004, \mnras, 350, 1174 
\bibitem{} Brunetti, G., Venturi, T., Dallacasa, D., et al.\ 2007, \apjl, 670, L5 
\bibitem{} Brunetti, G., Lazarian A., 2007, \mnras, 378, 245
\bibitem{} Brunetti, G., Cassano, R., Dolag, K., Setti, G.\ 2009, \aap, 507, 661 
\bibitem{} Brunetti, G., \& Lazarian, A.\ 2011a, \mnras, 410, 127
\bibitem{} Brunetti, G., \& Lazarian, A.\ 2011b, \mnras, 412, 817 
\bibitem{} Brunetti, G., Blasi, P., Reimer, O., et al.\ 2012, \mnras, 426, 956 
\bibitem{} Brunetti, G., Rudnick, L., Cassano, R., et al.\ 2013, \aap, 558, A52 
\bibitem{} Brunetti, G., \& Jones, T.~W.\ 2014, International Journal of Modern Physics D, 23, 1430007
\bibitem{} Carilli, C.L., Taylor, G.B.\ 2002, \araa, 40, 319
\bibitem{} Cassano, R., \& Brunetti, G.\ 2005, \mnras, 357, 1313
\bibitem{} Cassano, R., Brunetti, G., R\"ottgering, H.J.A, Br\"uggen, M.\ 2010, \aap, 509, A68
\bibitem{} Cassano, R., Ettori, S., Brunetti, G., et al.\ 2013, \apj, 777, 141 
\bibitem{} Chandran, B.~D.~G.\ 2000, Physical Review Letters, 85, 4656 
\bibitem{} Cuciti, V., Cassano, R., Brunetti, G., Dallacasa, D., Kale. R., Ettori, S., Venturi, T.\ 2015, arXiv:1506.03209
\bibitem{} Daughton, W., Roytershteyn, V., Karimabadi, H., et al.\ 2011, Nature Physics, 7, 539 
\bibitem{} de Gouveia dal Pino, E.~M., \& Lazarian, A.\ 2005, \aap, 441, 845
\bibitem{} Donnert, J., Dolag, K., Brunetti, G., \& Cassano, R.\ 2013, \mnras, 429, 3564
\bibitem{} Donnert, J., Brunetti, G., \ 2014, \mnras, 443, 3564
\bibitem{} Drake, J.~F., Swisdak, M., Che, H., \& Shay, M.~A.\ 2006, \nat, 443, 553
\bibitem{} Drake, J.~F., Opher, M., Swisdak, M., \& Chamoun, J.~N.\ 2010, \apj, 709, 963 
\bibitem{} Drake, J.~F., Swisdak, M., \& Fermo, R.\ 2013, \apjl, 763, L5 
\bibitem{} Drury, L. O'C., 2012, \mnras, 422, 2474
\bibitem{} Eyink, G.~L.\ 2011, \pre, 83, 056405 
\bibitem{} Eyink, G., Vishniac, E., Lalescu, C., et al.\ 2013, \nat, 497, 466
\bibitem{} Eyink, G.~L.\ 2015, \apj, 807, 137
\bibitem{} Ettori, S., Pratt, G.W., de Plaa, J., et al.\ 2013, arXiv:1306.2322
\bibitem{} Feretti, L., Giovannini, G., Govoni, F., \& Murgia, M.\ 2012, \aapr, 20, 54
\bibitem{} Fisk, L.~A.\ 1976, \jgr, 81, 4633 
\bibitem{} Fujita, Y., Takizawa, M., \& Sarazin, C.~L.\ 2003, \apj, 584, 190 
\bibitem{} Fujita, Y., Takizawa, M., Yamazaki, R., Akamatsu, H., Ohno, H.\ 2015, \apj, 815, 116
\bibitem{} Giannios, D.\ 2013, \mnras, 431, 355
\bibitem{} Kadowaki, L.~H.~S., de Gouveia Dal Pino, E.~M., \& Singh, C.~B.\ 2015, \apj, 802, 113 
\bibitem{} Karimabadi, H., \& Lazarian, A.\ 2013, Physics of Plasmas, 20, 112102 
\bibitem{} Khiali, B., de Gouveia Dal Pino, E.~M., \& del Valle, M.~V.\ 2015, \mnras, 449, 34 
\bibitem{} Kitayama, T., Bauts, M., Markevitch, M., et al.\ 2014, arXiv:1412.1176
\bibitem{} Kowal, G., Lazarian, A., Vishniac, E.~T., \& Otmianowska-Mazur, K.\ 2009, \apj, 700, 63 
\bibitem{} Kowal, G., de Gouveia Dal Pino, E.~M., \& Lazarian, A.\ 2011, \apj, 735, 102
\bibitem{} Kowal, G., de Gouveia Dal Pino, E.~M., \& Lazarian, A.\ 2012, Physical Review Letters, 108, 241102
\bibitem{} Lalescu, C.~C., Shi, Y.-K., Eyink, G.~L., et al.\ 2015, Physical Review Letters, 115, 025001
\bibitem{} Lazarian A., Vishniac E.T., 1999, \apj, 517, 700 (LV99)
\bibitem{} Lazarian, A., \& Beresnyak, A.\ 2006, \mnras, 373, 1195 
\bibitem{} Lazarian, A., \& Opher, M.\ 2009, \apj, 703, 8
\bibitem{} Lazarian, A., \& Desiati, P.\ 2010, \apj, 722, 188 
\bibitem{} Lazarian, A., \& Brunetti, G.\ 2011, \memsai, 82, 636 
\bibitem{} Lazarian, A., Eyink, G.~L., Vishniac, E.~T., \& Kowal, G.\ 2015, ASSL, 407, 311
\bibitem{} Le Roux, J.A, Zank G.P, Webb G.M., Khabarova O.\ 2015, \apj, 801, 112
\bibitem{} Liang, H., Hunstead, R.~W., Birkinshaw, M., \& Andreani, P.\ 2000, \apj, 544, 686 
\bibitem{} Longair, M.~S.\ 2011, {\it High Energy Astrophysics}, Cambridge, UK: Cambridge University Press, 2011
\bibitem{} Loureiro, N.~F., Schekochihin, A.~A., \& Cowley, S.~C.\ 2007, Physics of Plasmas, 14, 100703 
\bibitem{} Lyubarsky, Y.~E.\ 2003, \mnras, 345, 153 
\bibitem{} Lyutikov, M., \& Blandford, R.\ 2003, arXiv:astro-ph/0312347 
\bibitem{} Macario, G., Venturi, T., Intema, H.~T., et al.\ 2013, \aap, 551, A141 
\bibitem{} Makwana, K.D, Zhdankin V., Li, H., Daughton, W., Cattaneo, F. \ 2015, Physics of Plasmas, 22, 042902
\bibitem{} Marchegiani, P., \& Colafrancesco, S.\ 2015, \mnras, 452, 1328 
\bibitem{} Mertsch , P. \ 2011, \jcap, 12, 010
\bibitem{} Miniati, F.\ 2014, \apj, 782, 21 
\bibitem{} Miniati, F.\ 2015, \apj, 800, 60
\bibitem{} Miniati, F., \& Beresnyak, A.\ 2015, \nat, 523, 59
\bibitem{} Ng, J., Huang, Y.-M., Hakim, A., et al. \ 2015, arXiv:1511.00741
\bibitem{} Ohno, H., Takizawa, M., \& Shibata, S.\ 2002, \apj, 577, 658 
\bibitem{} Oishi, J.~S., Mac Low, M.-M., Collins, D.~C., \& Tamura, M.\ 2015, \apjl, 806, L12 
\bibitem{} Owen, F.~N., Rudnick, L., Eilek, J., et al.\ 2014, \apj, 794, 24 
\bibitem{} Parker, E.~N.\ 1957, \jgr, 62, 509
\bibitem{} Petrosian, V.\ 2001, \apj, 557, 560
\bibitem{} Petrosian, V., \& East, W.~E.\ 2008, \apj, 682, 175 
\bibitem{} Petschek, H.~E.\ 1964, NASA Special Publication, 50, 425 
\bibitem{} Pinzke, A., Oh, S.~P., \& Pfrommer, C.\ 2015, arXiv:1503.07870 
\bibitem{} Porter, D.~H., Jones, T.~W., \& Ryu, D.\ 2015, \apj, 810, 93 
\bibitem{} Ryu, D., Kang, H., Cho, J., \& Das, S.\ 2008, Science, 320, 909 
\bibitem{} Santos-Lima, R., de Gouveia Dal Pino, E.~M., Kowal, G., et al.\ 2014, \apj, 781, 84 
\bibitem{} Schlickeiser, R., {\it Cosmic ray astrophysics}, Springer (2002)
\bibitem{} Schlickeiser, R.\ 1984, \aap, 136, 227  
\bibitem{} Schlickeiser, R., \& Miller, J.~A.\ 1998, \apj, 492, 352 
\bibitem{} Shay, M.~A., Drake, J.~F., Denton, R.~E., \& Biskamp, D.\ 1998, \jgr, 103, 9165 
\bibitem{} Shimwell, T.~W., Brown, S., Feain, I.~J., et al.\ 2014, \mnras, 440, 2901 
\bibitem{} Sironi, L., \& Spitkovsky, A.\ 2014, \apjl, 783, L21 
\bibitem{} Sweet, P.~A.\ 1958, Electromagnetic Phenomena in Cosmical Physics, 6, 123 
\bibitem{} Uzdensky, D.~A., Loureiro, N.~F., \& Schekochihin, A.~A.\ 2010, Physical Review Letters, 105, 235002 
\bibitem{} Vazza F., Brunetti G., Gheller C., Brunino C., Br\"uggen M.\ 2011, \aap, 529, A17
\bibitem{} Venturi, T.\ 2011, \memsai, 82, 499 
\bibitem{} Yan, H., \& Lazarian, A.\ 2004, \apj, 614, 757 
\bibitem{} Yan, H., \& Lazarian, A.\ 2008, \apj, 673, 942
\bibitem{} Zandanel, F., \& Ando, S.\ 2014, \mnras, 440, 663 
\bibitem{} Zhang, B., \& Yan, H.\ 2011, \apj, 726, 90 
\bibitem{} Zank, G.~P., le Roux, J.~A., Webb, G.~M., Dosch, A., \& Khabarova, O.\ 2014, \apj, 797, 28 
\bibitem{} ZuHone, J.~A., Markevitch, M., Brunetti, G., \& Giacintucci, S.\ 2013, \apj, 762, 78 
\end{thebibliography}
\end{document}